\documentclass[preprint,12pt]{elsarticle}
\usepackage{amssymb}
\usepackage{latexsym}
\usepackage{graphicx}
\usepackage{epstopdf}
\usepackage{amsthm}
\usepackage{fixltx2e}
\MakeRobust{\bf}

\journal{Physics Letters A}

\begin{document}

\begin{frontmatter}


\title{The choice of a V-type system with the probe field in the subluminal or superluminal regimen} 

\author{O. Budriga\corref{cor1}}

\address{Laser Department, National Institute for Laser, Plasma and Radiation Physics, P.O. Box MG-36, 077125, Magurele, Romania}

\ead{olimpia.budriga@inflpr.ro}

\cortext[cor1]{Tel./Fax:+40214574467}

\begin{abstract}
We study the influence of the angle between the two dipole transition moments on the group velocity for two kinds of three-level V-type systems with spontaneously generated coherence and incoherent pumping. The group velocity of the probe field for both kinds of the V-type systems alternates between the subluminal or superluminal regimen. We can choose a real three-level V-type system from atoms or molecules to exhibit subluminal or superluminal light propagation by a proper geometry of the probe, coupling and incoherent pumping fields, which depends on the angle between the two dipole transition moments.

\end{abstract}


\end{frontmatter}

\section{Introduction}
It is well known that the group velocity of a laser can be subluminal or superluminal in atomic gases \cite{hau,liu,wang,kim,bae} and solid praseodymium doped in Y\textsubscript{2}SiO\textsubscript{5} \cite{longdell}. The control of the group velocity can be accomplished through the electromagnetically induced transparency and electromagnetically induced amplification \cite{kim,bae,bigelow,agarwal,han,joshi}. The spontaneously generated coherence (SGC) has a major influence on the regimen of the group velocity. The effects of the SGC on the change of the group velocity from subluminal to sperluminal was studied for three-level Lambda systems without \cite{mahmoudi} and with \cite{dastidar} incoherent pumping. The three-level V-type system is another model of system of interest in which it can be achieved a probe field group velocity lower or higher than the light speed {\it c} ($c\cong3\cdot10^8$ m/s). There are two kinds of the three-level V-type system which was shown in our work \cite{OliLaserPhysics2014}. In the three-level V-type system of the first kind one field acts on only one transition. The relative phase effects on the group velocity in a V-type system of the first kind with spontaneously generated coherence and without incoherent pumping leads to the subluminal and superluminal group velocity \cite{han}. The same change from subluminal to superluminal regimen of the group velocity of the probe field can be obtained by variation of the incoherent pumping rate and the SGC parameter, which depends on the angle between the two induced dipole moments, in the V-type system of the first kind with incoherent pumping \cite{guo}. In the three-level V-type system of the second kind all three fields, the probe, the coupling and the incoherent pumping (if is considered) fields drive both optical allowed transitions. The group velocity can pass from the subluminal to superluminal regimen and viceversa by the modification of the relative phase between the probe and coupling fields, too, in the case of this type of system \cite{arbiv}. In a recent paper sent to publication \cite{OliLaserPhysics2014} we found that in both kinds of the three-level V-type systems the proper choice of the relative phase between the probe and coupling fields and the incoherent pumping rate leads to the subluminal or superluminal propagation of the probe field. 

Another parameter which can impinge on the group velocity in both kinds of three-level V-type systems with SGC is the angle between the two dipole transition moments, denoted with $\theta$. In the case of the three-level V-type system of the first kind the knob effect of the parameter which is direct proportional with $\theta$ was relieved in connection with the incoherent pumping rate, but without the dependence on the relative phase \cite{guo}. Our goal is to obtain the dependence of the group velocity on the the angle between the two dipole transition moments and the relative phase for the both kinds of the three-level V-type systems described in our previous paper \cite{OliLaserPhysics2014}. We presented a real V-type system of the second kind from the LiH molecule in our previous work \cite{OliOptCommun2014}. In this case the angle $\theta = \pi$ and we showed that it can exhibit subluminal or superluminal group velocity \cite{OliLaserPhysics2014}. We will compare these results with those obtained from the general formulas in angle $\theta$ for the V-type system of the second kind in the particular case of antiparallel dipole moments when $\theta = \pi$.    

A brief presentation of both three-level V-type systems and the formulas of the group index will be given in Section \ref{sec:2}. The Subsection \ref{subsec2.1} is devoted to the three-level V-type system of the first kind and the Subsection \ref{subsec2.2} to the three-level V-type system of the second kind. Numerical results of the dependence on the $\theta$ angle of the index group for both kinds of V-type systems are shown in Section \ref{sec:3}. We conclude with some remarks in the last section.

\section{Theory}
\label{sec:2}
We divided the three-level V-type systems with spontaneously generated coherence and incoherent pumping in two kinds \cite{OliLaserPhysics2014}. Below we will describe them and the index group formulas depending on the density matrix elements and the other parameters of the V-type system with incoherent pumping.

\subsection{The V-type system of the first kind}
\label{subsec2.1}
\begin{figure*}
  \includegraphics{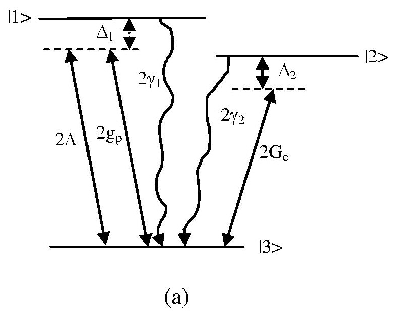}
   \includegraphics{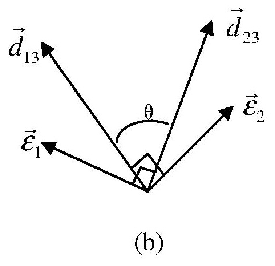}
\caption{(a) The three-level V-type system with two near-degenerated excited states $|1\rangle $, $|2\rangle $ and a ground level $|3\rangle $. (b) The electric dipole transition moments are chosen so that one field acts on only one transition. }
\label{fig1}      
\end{figure*}

The closed three-level V-type system of the first kind is the most known V-type system. It consists of two closely-lying excited states $\vert1\rangle$ and $\vert2\rangle$ and a ground state $\vert3\rangle$, as shown in Figure \ref{fig1}(a). The rates of spontaneous emission from levels $\vert1\rangle $ and $\vert2\rangle $ to ground level $|3\rangle $ are denoted by $2\gamma_{1}$ and $2\gamma_{2}$, respectively. The transition $|2\rangle \leftrightarrow |3\rangle $ with frequency $\omega_{23}$ are driven by a coupling coherent field (${\bf{E}}_{2}={\bf \epsilon}_{2}e^{-i\omega_{2}t}+c.c.$) with the Rabi frequency $2G_{c}=2{\bf \epsilon}_{2}\cdot{\bf{d}}_{23}/\hbar$, where ${\bf{d}}_{23}$ is the transition dipole moment. Between the levels $\vert1\rangle $ and $|3\rangle $ is applied a probe field (${\bf{E}}_{1}={\bf{\epsilon}}_{1}e^{-i\omega_{1}t}+c.c.$) with the Rabi frequency $2g_{p}=2{\bf{\epsilon}}_{1}\cdot{\bf{d}}_{13}/\hbar$. The transition $|1\rangle\leftrightarrow |3\rangle$ with the frequency $\omega_{13}$ and dipole moment ${\bf{d}}_{13}$ is pumped with a rate $2\Lambda$ by an incoherent field. The detunings of the probe field and the coupling field are $\Delta_{1} = \omega_{13} - \omega_{1}$ and $\Delta_{2} = \omega_{23} - \omega_{2}$, respectively.

The density matrix formalism was used to describe this system by the statistical view. The density matrix equations system (2) from \cite{OliLaserPhysics2014} contains two important parameters related to the SGC: the parameter $\eta$ and the relative phase between the probe and coupling field phases $\phi$. The parameter $\eta$ describes the quantum interference between spontaneous emission from excited levels $|1\rangle$ and $|2\rangle$ to ground level $|3\rangle$ and depends on $\theta$, the angle between the two dipole momentum ${\bf{d}}_{13}$ and ${\bf{d}}_{23}$, $\eta=\eta_{0}\sqrt{\gamma_{1}\gamma_{2}}\cos\theta$, where $\eta_{0}\equiv(\omega_{23}/\omega_{13})^{3/2}$ \cite{cardimona}. The SGC effects are important only for small energy spacing between the two excited levels \cite{Agarwal}, and for high energy spacing such effects will disappear \cite{Paspalakis}. As the excited levels $|1\rangle$ and $|2\rangle$ are near-degenerated, then $\omega_{23}\approx\omega_{13}$ and $\eta_{0}\approx1$. The condition to have spontaneously generated coherence is $\eta\neq0$, which means that we must have a nonorthogonal dipol momentum of the two transitions. Therefore we choose the dipole momentum so that one field acts on one transition (${\bf{d}}_{13}\perp{\bf{\epsilon}}_{2}$, ${\bf{d}}_{23}\perp{\bf{\epsilon}}_{1}$) as it can be shown in Figure \ref{fig1}(b). Rabi frequencies are connected to the angle $\theta$ by the relations $2g_{p}=2|{\bf{E}}_{1}||{\bf{d}}_{13}|\sin\theta/\hbar$ and $2G_{c}=2|{\bf{E}}_{2}||{\bf{d}}_{23}|\sin\theta/\hbar$. The influence of the angle $\theta$ on the probe coefficient absorption and refractive index was evidenced in our paper \cite{Oli2013}. The nonorthogonality of the dipolar momentum can be achieved from the mixing of the levels arising from internal \cite{physrevlet77} or external fields \cite{hakuta, faist, berman, patnaik, optcomun179}. 

The formula of the group index was deduced by us in our previous work \cite{OliLaserPhysics2014}
 \begin{eqnarray}
\label{ec0}
n_{\rm g}-1=\frac{N d_{31}^2}{\hbar\epsilon_{0}g}\lbrack Re\tilde\rho_{31}-\omega_{1}\frac{d Re\tilde\rho_{31}}{d \Delta_{1}}\rbrack.
\end{eqnarray}
with the derivative of the real part of the density matrix element $\tilde{\rho}_{31}$, $Re\tilde{\rho}_{31}$, with respect to the probe detuning $\Delta_{1}$ given in the Appendix of \cite{OliLaserPhysics2014}. This relation will be used in our numerical calculations.

\subsection{The V-type system of the second kind}
\label{subsec2.2}
The three-level V-type system of the second kind is the system from the Figure \ref{fig1}(a) but where the probe, the coupling and the incoherent pumping fields act on both transitions. The density matrix equations system differs from the system of linear equations (1) from \cite{OliLaserPhysics2014} for the V-type system of the first kind by the appearance of the term $2\Lambda\tilde{\rho}_{33}$ in the second equation and can be written as  
$$-2\gamma_{1}\tilde{\rho}_{11}+2\Lambda\tilde{\rho}_{33}-\eta e^{i\phi}\tilde\rho_{12}-\eta e^{-i\phi}\tilde\rho_{21}+i(g+G')\tilde\rho_{31}-i(g+G')\tilde\rho_{13}=0$$
$$-2\gamma_{2}\tilde{\rho}_{22}+2\Lambda\tilde{\rho}_{33}-\eta e^{i\phi}\tilde{\rho}_{12}-\eta e^{-i\phi}\tilde{\rho}_{21}+i(G+g')\tilde{\rho}_{32}-i(G+g')\tilde{\rho}_{23}=0$$
$$-(\gamma_{1}+2\Lambda+i\Delta_{1})\tilde{\rho}_{13}-\eta e^{-i\phi}\tilde{\rho}_{23}+i(g+G')(\tilde{\rho}_{33}-\tilde{\rho}_{11})-i(G+g')\tilde{\rho}_{12}=0$$
$$-(\gamma_{2}+2\Lambda-i\Delta_{2})\tilde{\rho}_{23}-\eta e^{i\phi}\tilde{\rho}_{13}- \nonumber\\
i(g+G')\tilde{\rho}_{21}+i(G+g')(\tilde{\rho}_{33}-\tilde{\rho}_{22})=0$$
$$-[\gamma_{1}+\gamma_{2}+i(\Delta_{1}-\Delta_{2})]\tilde{\rho}_{12}-\eta e^{-i\phi}(\tilde{\rho}_{11}+\tilde{\rho}_{22})+$$
\begin{equation}
\label{ec1}
i(g+G')\tilde{\rho}_{32}-i(G+g')\tilde{\rho}_{13}=0
\end{equation}
with {\it g, g', G, G'} the real Rabi frequencies. The other equations have the same terms as (1) from \cite{OliLaserPhysics2014}, where the $\Lambda$, $g$ and $G$ are replaced by $2\Lambda$, $g+G'$ and $G+g'$, respectively. The dependence on the angle between the dipole transition moments ${\bf{d}}_{13}$ and ${\bf{d}}_{23}$, $\theta$ is contained in the same paramater $\eta$ as in the case of the V-type system of the first kind, but additionally it appears in the Rabi frequencies formulas as following
\begin{eqnarray}
\label{ec2}
2g &=& 2E_{1}d_{13}/\hbar\\
2g' &=& 2E_{1}d_{23}\cos{\theta}/\hbar\\
2G &=& 2E_{2}d_{23}/\hbar\\
2G' &=& 2E_{2}d_{13}\cos{\theta}/\hbar,
\end{eqnarray}
if we choose the probe field ${\bf{E}}_{1}$ parallel with the dipole transition moment ${\bf{d}}_{13}$ and the coupling field ${\bf{E}}_{2}$ parallel with the dipole transition moment ${\bf{d}}_{23}$.  

Using the relations (10) and (11) from \cite{OliLaserPhysics2014} the electric susceptibility of the medium $\chi_{e}$ becomes
\begin{eqnarray}
\label{ec3} 
\chi_{e} = \frac{2N(\tilde{\rho}_{13}{\bf{d}}_{31}+\tilde{\rho}_{23}{\bf{d}}_{32})}{\epsilon_{0}{\bf{E}}_{1}}.
\end{eqnarray}
The choise of the dipole transition moments from above gives us the formula of the electric susceptibility of the medium as
\begin{eqnarray}
\label{ec4} 
\chi_{e} = \frac{2N(\tilde{\rho}_{13}d_{13}+\tilde{\rho}_{23}d_{23}\cos{\theta})}{\epsilon_{0}E_{1}}.
\end{eqnarray}
The quantity that we will study is the group index $n_{g}-1$, which is related to the real part of electric susceptibility and is
\begin{eqnarray}
\label{ec5} 
n_{g}-1 = \frac{Nd_{13}}{\epsilon_{0}g\hbar}\lbrace\lbrack Re\tilde{\rho}_{13} - (\omega_{13}-\Delta_{1})\frac{dRe\tilde{\rho}_{13}}{d\Delta_{1}}\rbrack d_{13} +\nonumber\\ \lbrack Re\tilde{\rho}_{23} - (\omega_{13}-\Delta_{1})\frac{dRe\tilde{\rho}_{23}}{d\Delta_{1}}\rbrack d_{23}\cos{\theta}\rbrace.
\end{eqnarray} 
The difference between the above relation and the expression (13) of the group index from our work \cite{OliLaserPhysics2014} is the presence of the $\cos{\theta}$ in the second term of the above relation. The real V-type system of the second kind from LiH molecule studied by us in \cite{OliLaserPhysics2014} is a particularly case when the angle between the two dipole transition moments $\theta$ is $\pi$.

\section{Results}
\label{sec:3}
\begin{figure}
 \resizebox{0.75\columnwidth}{!}{
  \includegraphics{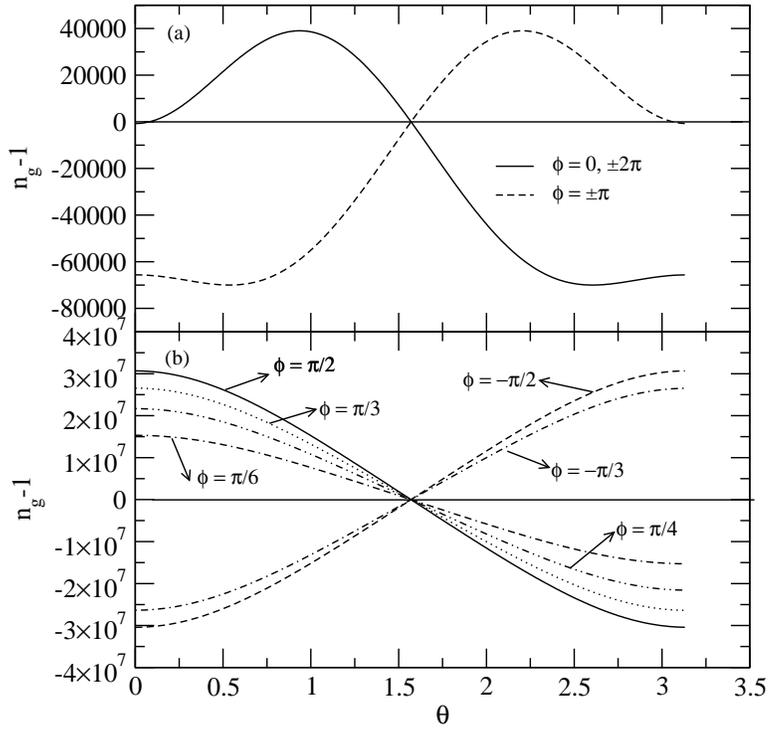}
  }
\caption{The group index $n_{g}-1$ as a function of the angle between the two dipole transition moments, $\theta$, in the case of the V-type system of the first kind for some values of the relative phase $\phi$: (a) 0, $\pm2\pi$, $\pm\pi$ and (b) $\pm\pi/2, \pm\pi/3, \pi/4, \pi/6$. The other parameters of the system are: $\gamma_{1} = 2.5\cdot10^{4}$ Hz, $\gamma_{2} = 0.33 \gamma_{1}$, $\Delta_{2} = 0$, $\omega_{1} = 10^{10} \gamma_{1}$, $\Lambda = \gamma_{1}$, $g = 0.033 \gamma_{1}$, $G = 60,76 \gamma_{1}$, $N = 10^{18}$ molecules/m\textsuperscript{3}.}
\label{fig2}      
\end{figure}

In Figure \ref{fig2} can be shown that for the three-level V-type system of the first kind, for each value of the angle $\theta$ between the two dipole transition moments ${\bf{d}}_{13}$ and ${\bf{d}}_{23}$ one can choose the value of the relative phase $\phi$ to have a subluminal ($n_{g}-1 > 0$) or superluminal ($n_{g}-1 < 0$) probe field group velocity. The values of the group index are lower with three order of magnitude for the relative phase $\phi = 0, \pm2\pi, \pm\pi$ (see Figure \ref{fig2} (a)) than the relative phase equal with $\pm\pi/2, \pm\pi/3, \pi/4, \pi/6$ (see Figure \ref{fig2} (b)). We considered the parameters of the system as: $\gamma_{1} = 2.5\cdot10^{4}$ Hz, $\gamma_{2} = 0.33 \gamma_{1}$, $\Delta_{2} = 0$, $\omega_{1} = 10^{10} \gamma_{1}$, $\Lambda = \gamma_{1}$, $g = 0.033 \gamma_{1}$, $G = 60,76 \gamma_{1}$, $N = 10^{18}$ molecules/m\textsuperscript{3}. Another behaviour of the group index we can see from the Figure \ref{fig2} (a) is that it is periodically with the relative phase $\phi$, with the period $2\pi$.    

\begin{figure}
 \resizebox{0.75\columnwidth}{!}{
  \includegraphics{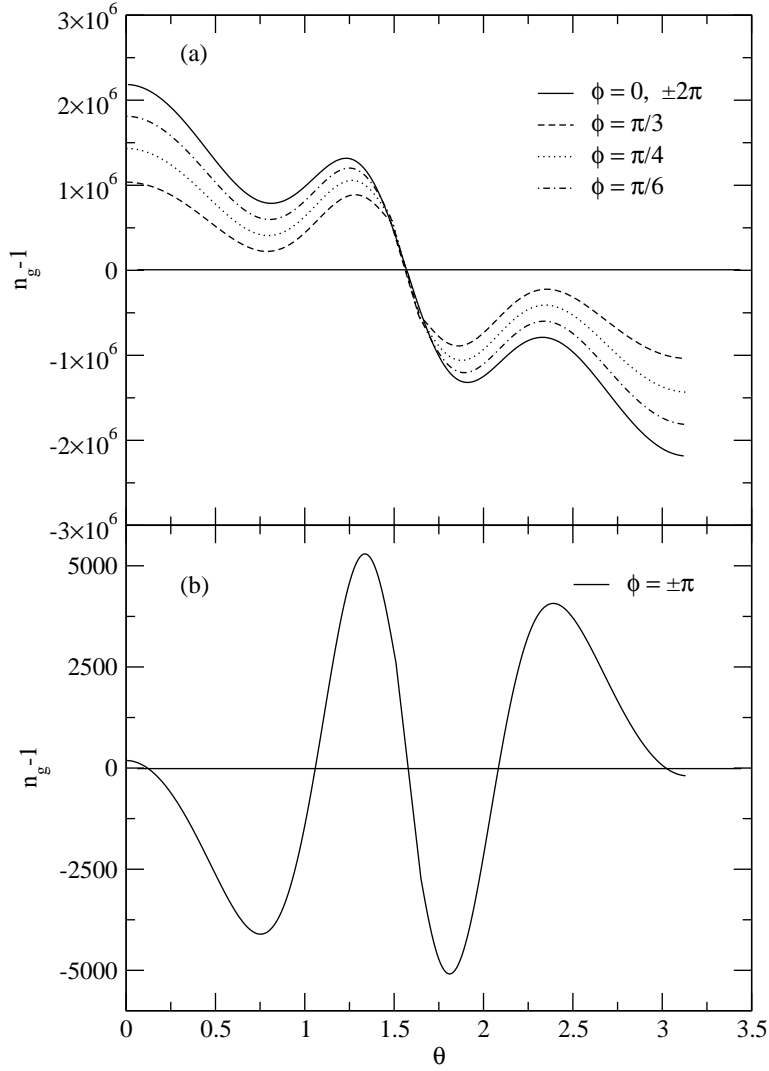}
  }
\caption{The group index $n_{g}-1$ as a function of $\theta$ for the V-type system of the second kind. The relative phase $\phi$ is: (a) 0, $\pm2\pi$, $\pi/3$, $\pi/4$, $\pi/6$ and (b) $\pm\pi$. The other parameters of the system are: $\gamma_{1} = 2.475\cdot10^{4}$ Hz, $\gamma_{2} = 0.33 \gamma_{1}$, $\Delta_{2} = 0$, $\omega_{1} = 10^{10} \gamma_{1}$, $\Lambda = \gamma_{1}$, $g = 0.033 \gamma_{1}$, $G = 235.35 \gamma_{1}$, $\omega_{13} = 2\cdot10^{11}\gamma_{1}$, $N = 10^{18}$ molecules/m\textsuperscript{3}.}
\label{fig3}      
\end{figure}

The Figure \ref{fig3} is dedicated to the three-level V-type system of the second kind. Numerical results of the group index $n_{g}-1$ from equation (\ref{ec5}) are plotted for few values of the relative phase $\phi$ in the range $[-2\pi,2\pi]$ and the angle $\theta\in[0,\pi]$. The system which we choose is the real LiH molecule V-type system of the second kind which we have shown in \cite{OliLaserPhysics2014} with the parameters $\gamma_{1} = 2.475\cdot10^{4}$ Hz, $\gamma_{2} = 0.33 \gamma_{1}$, $\Delta_{2} = 0$, $\omega_{1} = 10^{10} \gamma_{1}$, $\Lambda = \gamma_{1}$, $g = 0.033 \gamma_{1}$, $G = 235.35 \gamma_{1}$, $\omega_{13} = 2\cdot10^{11}\gamma_{1}$, $N = 10^{18}$  molecules/m\textsuperscript{3}. The Rabi frequencies $g'$ and $G'$ are given by the relations (4) and (6), respectively.     
For the angle $\theta\in[0,\pi/2)$ and every relative phase $\phi\ne\pm\pi$ the group index $n_{g}-1$ has only positive values which means the probe field group velocity is subluminal as can be seen in Figure \ref{fig3}(a). The group index $n_{g}-1 < 0$ when $\theta\in(\pi/2,\pi]$ and relative phase $\phi\ne\pm\pi$, then the probe field group velocity is superluminal. The group index is null when the angle $\theta=\pi/2$. This is an expected result, because for $\theta=\pi/2$ there is no effect of the spontaneously generated coherence. In Figure \ref{fig3} (b) can be seen the behaviour of the group index in the case of the relative phase $\phi=\pm\pi$, which is different and its values are lower with three order of magnitude than in the other case from Figure \ref{fig3} (a). It is kept the periodicity with relative phase $\phi$ with the period $2\pi$. The straightforward calculations of the group index $n_{g}-1$ using the relation (\ref{ec5}) and the numerical solutions of the equation system (\ref{ec1}) in the case of the real V-type system of the second kind from the LiH molecule \cite{OliLaserPhysics2014} lead us to the same graphic as in the Figure 7(b) for the incoherent pumping rate $\Lambda = \gamma_{1}$.

\section{Conclusions}
\label{sec:4}
We studied the influence of the angle between the two dipole transition moments, denoted $\theta$, on the group velocity of the probe field for both kinds three-level V-type systems. We have shown that the group velocity of the probe field in the case of the V-type system of the first kind can be subluminal or superluminal for each value of the angle $\theta$ if we change the value of the relative phase $\phi$. The V-type system of the second kind is different. The group velocity of the probe field is lower than the light velocity for every value of the relative phase $\phi\ne\pm\pi$ and the angle $\theta\in[0,\pi/2)$. For an angle $\theta\in(\pi/2,\pi]$ the group velocity of the probe field is higher than the light speed not depending on the relative phase $\phi\ne\pm\pi$. The real system of the LiH molecule of the V-type system of the second kind confirms our previous results for the particularly value of the angle $\theta = \pi$.

\section{Acknowledgements}
The author thanks the support of Ministry of Education and Research, Romania (program Laplas 3, PN 09 39).


\begin{thebibliography}{00}
\bibitem{hau} L. V. Hau, S. E. Harris, Z. Dutton and C. H. Behroozi, Nature (London) 397 (1999) 594.
\bibitem{liu} C. Liu, Z. Dutton, C. H. Behroozi and L. V. Hau, Nature (London) 409 (2001) 490.
\bibitem{wang} L. J. Wang, A. Kuzmich and A. Dogariu, Nature (London) 406 (2000) 277.
\bibitem{kim} K. Kim, H. S. Moon, C. Lee, S. K. Kim and J. B. Kim, Phys. Rev. A 68 (2003) 013810.
\bibitem{bae}I.-H. Bae and H. S. Moon, Phys. Rev. A 83 (2011) 053806.
\bibitem{longdell} J. J. Longdell, E. Fraval, M. J. Sellars and N. B. Manson, Phys. Rev. Lett. 95 (2005) 063601. 
\bibitem{bigelow} M. S. Bigelow, N. N. Lepeshkin and R. W. Boyd, Science 301 (2003) 200.
\bibitem{agarwal}G. S. Agarwal, T. N. Dey and S. Menon, Phys. Rev. A 64 (2001) 053809.
\bibitem{han}D. Han, H. Guo, Y. Bai and H. Sun, Phys. Lett. A 334 (2005) 243.
\bibitem{joshi} A. Joshi, S. S. Hassan and M. Xiao, Phys. Rev. A 72 (2005) 055803.
\bibitem{mahmoudi} M. Mahmoudi, M. Sahrai and H. Tajalli, J. Phys. B: At. Mol. Opt. Phys. 39 (2006) 1825; Mahmoudi M, M. Mahmoudi, M. Sahrai and H. Tajalli, Phys. Lett. A 357 (2006) 66.
\bibitem{dastidar} S. Dutta and R. Dastidar, J. Phys. B: At. Mol. Opt. Phys. 40 (2007) 4287.
\bibitem{OliLaserPhysics2014}O. Budriga, arXiv:1406.5288 (2014). 
\bibitem{guo}Y. Bai, H. Guo, D. Han and H. Sun, Phys. Lett. A 357 (2005) 66.
\bibitem{arbiv} D. Bortman-Arbiv, A. D. Wilson-Gordon and H. Friedmann, Phys. Rev. A 63 (2001) 043818.
\bibitem{OliOptCommun2014}O. Budriga, Opt. Commun. 328 (2014) 77. 
\bibitem{cardimona}D. A. Cardimona, M. G. Raymer, C. R. Stroud Jr, J. Phys. B: At. Mol. Phys. 15 (1982) 55.   
\bibitem{Agarwal} S. Menon and G. S. Agarwal, Phys. Rev. A 57 (1998) 4014.
\bibitem{Paspalakis} E. Paspalakis, S. Q. Gong and P. L. Knight, J. Mod. Opt. 45 (1998) 2433.
\bibitem{Oli2013} O. Budriga, Phys. Scr. T153 (2013) 014007.
\bibitem{physrevlet77} H. R. Xia, C. Y. Ye and S. Y. Zhu, Phys. Rev. Lett. 77 (1996) 1032.   
\bibitem{hakuta}K. Hakuta, L. Marmet and B. P. Stoicheff, Phys. Rev. Lett. 66 (1991) 596.   
\bibitem{faist}J. Faist, F. Capasso, C. Sirtori, K. V. West and L. N. Pfieffer, Nature (London) 390 (1997) 589.   
\bibitem{berman}P. R. Berman, Phys. Rev. A 58 (1998) 4886.    
\bibitem{patnaik}A. K. Patnaik and G. S. Agarwal, Phys. Rev. A 59 (1999) 3015.   
\bibitem{optcomun179}P. Zhou and S. Swain, Opt. Commun. 179 (2000) 267.   
 


\end{thebibliography}
\end{document}